# Proton Collective Quantum Tunneling Induces Anomalous Thermal Conductivity of Ice under Pressure

Yufeng Wang[1,2]*, Ripeng Luo[2]*, Jian Chen[3], Xuefeng Zhou[3], Shanmin Wang[3], Junqiao Wu[4,5], Feiyu Kang[1,2], Kuang Yu[1,2]†, Bo Sun[1,2]†

[1]*Tsinghua Shenzhen International Graduate School, Tsinghua University, Shenzhen 518055, China.*

[2]*Tsinghua-Berkeley Shenzhen Institute, Tsinghua University, Shenzhen 518055, China.*

[3]*Department of Physics and Academy for Advanced Interdisciplinary Studies, Southern University of Science and Technology, Shenzhen 518055, China.*

[4]*Department of Materials Science and Engineering, University of California, Berkeley, CA 94720, USA.*

[5]*Materials Sciences Division, Lawrence Berkeley National Laboratory, Berkeley, CA 94720, USA.*

\* These authors contributed equally to this work

† e-mails: sun.bo@sz.tsinghua.edu.cn, yu.kuang@sz.tsinghua.edu.cn

**Proton tunneling is believed to be non-local in ice but has never been shown experimentally. Here we measured thermal conductivity of ice under pressure up to 50 GPa and found it to increase with pressure until 20 GPa but decrease at higher pressures. We attribute this anomalous drop of thermal conductivity to the collective tunneling of protons at high pressures, supported by large-scale quantum molecular dynamics simulations. The collective tunneling loops span several picoseconds in time and are as large as nanometers in space, which match the phonon periods and wavelengths, leading to strong phonon scattering at high pressures. Our results show direct evidence of collective quantum motion existing in high-pressure ice and provide a new perspective to understanding the coupling between phonon propagation and atomic tunneling.**

As one of the most important substances in the universe, ice challenges our understanding of its structures and properties, especially under extreme conditions, which are crucial in geological physics and environmental science[1–3]. In particular, the thermal conductivity ($\Lambda$) of pressurized ice plays an important role in the evolution and dynamics of icy planets, whose interior pressures range up to several hundred gigapascals (GPa)[4]. At such high pressures, the molecular structure of ice can be disrupted by hydrogen ion (proton) quantum tunneling, potentially affecting its thermal transport[5]. However, current data of $\Lambda$ of compressed ice is available to a maximum pressure of only 22 GPa[6], showing a monotonic increase with pressure consistent with the classical Leibfried-Schlömann (LS) equation, while the expected quantum effects in the dynamics of protons were obscure[7,8].

Meanwhile, ice has been an ideal model to study proton's quantum effects in the hydrogen bond network, which by themselves are long-standing and important topics in chemistry. Considering its simple molecular structure, ice features exceptionally complicated phase behavior under extreme conditions[9]. Specifically, at pressures higher than 2 GPa, the molecular crystal ice VII forms with an ordered oxygen (O) sublattice and a disordered hydrogen (H) sublattice. These sublattices are bounded by the "ice rule": each O is covalently bonded to two H atoms which are hydrogen-bonded to two other $H_2O$ molecules (Fig. 1). Each H resides in a double-well potential, allowing it to locally tunnel between neighboring O sites and shuffle the position of the H sublattice. Theoretical analysis indicates that such tunneling events of different H atoms are coupled to each other due to the instability of certain charged defects that disrupt the ice rule[7,10], leading to a nonlocal quantum motion that induces the phase transition to ice X, an atomic (as opposed to molecular) crystal with symmetric hydrogen bonds. The transition is fundamentally different from normal thermodynamic phase transitions[11,12] and is related to numerous anomalies observed in high-pressure ice[13–16]. However, previous experiments probing proton tunneling in high-pressure ice are based on spectroscopic techniques, such as infrared (IR)[17] and nuclear magnetic resonance (NMR)[18], which are sensitive only to the dynamics of single, isolated protons rather than collectively the entire hydrogen-bond network. As a result, the suspected nonlocal quantum motion has yet to be detected and the dynamic phase transition modulated by the tunneling is much less studied.

Here we report an anomalous, nonmonotonic pressure dependence of $\Lambda$ of both $H_2O$ and $D_2O$ ice up to 50 GPa, in stark contrast to what the classical LS equation predicts. We show that such anomaly is due to interactions between heat-carrying phonons and the proton tunneling, as the collective tunneling of multiple light atoms results in significant phonon scattering. Our work not only provides the widest range of thermal transport data for high-pressure ice, but also opens a new window to understanding its complicated phase diagram and phase transitions.

We measured $\Lambda$ of ice under pressure using an *in situ* time-domain thermoreflectance (TDTR) system (Extended Data Fig. 1). We put a thin sheet of metal-coated muscovite mica on the culet of diamond, then load liquid $H_2O$ or $D_2O$ into the diamond anvil cell (DAC). A 785 nm pump laser transmits through the diamond and ice to be focused onto the metal layer, where the laser beam is absorbed and converted into heat. The mica sheet is used as a thermal insulator, so that heat predominately flows into the surrounding ice rather than into the underlying diamond. A separate probe beam is used to monitor the temperature of the heated spot via the change of optical reflectance, and the resultant thermoreflectance signal is used to extract the thermal properties of ice. According to the sensitivity analysis of TDTR (Extended Data Fig. 2), the thickness ($h$) and volumetric heat capacity ($C_V$) of the metal transducer are the main parameters introducing measurement uncertainty. Therefore, two different metal transducers, Al and $Au_{0.95}Pd_{0.05}$ (AuPd), were used separately to ensure the reliability of experimental results. $\Lambda$ of ice in Fig. 2 was obtained by fitting to the experimental data with a bidirectional thermal diffusion model[6,19].

Below 20 GPa, $\Lambda$ measured in this work increases monotonically, in excellent agreement with previous work and the LS equation (Fig. 2 and Extended Data Fig. 3)[6]. However, beyond 20 GPa, we found that $\Lambda$ decreases until 40 GPa, then starts to level off and slightly increase between 40 and 50 GPa, forming a second turning point. We believe the measured thermal conductivity is intrinsic to ice due to the large crystal grain is much larger than phonon mean free path in ice, which is only a few nanometers (See the Supplementary Video and our calculation below). As an electrical insulator[20,21], $\Lambda$ of ice is completely dominated by lattice dynamics. It is well known that the lattice $\Lambda$ in a crystal is proportional to the group velocity of acoustic phonons, heat capacity, and phonon lifetimes. Our *in situ* time-domain stimulated Brillouin scattering data (Extended Data Fig. 4) indicates that the acoustic phonon velocity near $\Gamma$ point steadily increases over the entire pressure range we have studied. The heat capacity of ice VII is quite constant with a deviation of less than 30% in the entire pressure range (Extended Data Fig. 5). Therefore, the anomalous behavior of $\Lambda$ we observed in ice must result from a new scattering mechanism arising above 20 GPa that strongly reduces the phonon lifetimes. The quantum nature of the new scattering mechanism is clearly evidenced by comparing $\Lambda$ of $H_2O$ ice with that of $D_2O$ ice, which shows a similar trend but with the first turning point happening at about 10 GPa higher than $H_2O$ (Fig. 2b). Such a strong isotope effect points out the connection between the anomaly of $\Lambda$ and the quantum proton tunneling in the ice VII lattice, as heavier particles have a lower tunneling probability.

In order to understand the mechanism behind the anomalous behavior of $\Lambda$, large-scale equilibrium quantum molecular dynamics (MD) simulations were performed. A reactive machine learning (ML) potential was trained using the DeepMD-kit[22] at the Perdew–Burke-Ernzerhof (PBE) level of theory, and ring-polymer molecular dynamics (RPMD)[23] was used to propagate the dynamics. RPMD was proved to give a reliable description of both the thermodynamics and dynamics of short-living phonons in ice[24]. Both ice VII and X are isotropic, so the computed $\Lambda$ are averaged over three directions and plotted along the side with the experimental data in Fig. 2, showing quantitative agreement between them. We further present the integrated heat flux time correlation functions (TCFs) for $H_2O$ ice at different pressures (Extended Data Fig. 6), combined with the integrated TCFs from the O sublattice only. Comparing the total and the O integrated TCFs, it is clear that before 40 GPa, $\Lambda$ is primarily contributed by the O sublattice. Therefore, the heavy O phonons are the main heat carriers before 40 GPa, after which the H contribution rises as a consequence of the phonons propagating in the newly formed symmetrized H sublattice. The proton distribution profiles presented in Extended Data Fig. 7 support this picture, as the distinctive dual peaks start to merge at 40-50 GPa. Therefore, the second turning point of $\Lambda$ observed at about 40 GPa is directly related to the beginning of proton symmetrization and the phase transition to ice X.

With the O phonons identified as the main heat carrier, the first turning point at 20 GPa can be attributed to the scattering of the O phonons by the tunneling protons. We performed lattice dynamics analysis on the O sublattice, using the Green's function method[25,26]. The dispersion relation and the lifetimes of several selected O phonons are shown in Fig. 3 and Extended Data Fig. 8. At the vicinity of $\Gamma$ point, the group velocities gradually increase with pressure. It is consistent with the Brillouin scattering results (Extended Data Fig. 4) and also matches the common wisdom, as higher density usually leads to stronger force constants. However, the frequencies near the X point exhibit a reverse trend, downshifting between 20 and 40 GPa, which, as we will explain later, is a direct result of the charged defects caused by proton tunneling. Moreover, O phonon lifetimes in all points decrease steadily before 30-40 GPa, indicating that the O phonons are strongly scattered by the proton tunneling events (Fig. 3b).

To gain a deeper insight, we analyzed the tunneling pattern and found that the tunneling events are highly coupled and form a series of closed loops with varying sizes, as shown in Fig. 4. The time and spatial scales of these tunnel loops show strong pressure dependence. Between 20 GPa and 30 GPa, the protons still feature a far-separated dual peak distribution (Fig. 1 and Extended Data Fig. 7), indicating a thermally inaccessible barrier between the two sites. However, quantum tunnel loops spanning several picoseconds in time and as large as nanometers in space (Fig. 4c) can already be frequently observed. These values match the periods and wavelengths of the O phonons and cause strong phonon scattering, explaining the anomalous trend of $\Lambda$ in the corresponding pressure range. Above 30 GPa, the tunneling rate becomes too high and starts to be decoupled with the O phonons on the time scale. Especially, tunneling counts decrease in the low-frequency region below 2 THz (Fig. 4b) where phonons with large group velocities contribute to the majority of $\Lambda$. As a result, the lifetime of long-wavelength O phonons (*e.g.*, the one at (0 0 1/16)) begins to recover after 30 GPa (Fig. 3b).

Previous analysis indicated that in ice VII, protons tunnel in a synchronized fashion within six-member rings[10,27]. In this work, with a much larger simulation box, we find the collective tunneling happens over a larger spatial range and is not completely synchronized. At 300 K, many tunneling events are sequentially coupled via short-living ionic configurations, as indicated by previous studies[7]. The charged defects arising from the proton tunneling diffuse rapidly before being annihilated, forming large tunneling loops involving up to tens of molecules along the path. The O phonons with long wavelengths (~ nanometers) and periods (~ picoseconds) can be significantly scattered only by such collective tunneling loops, instead of by single tunneling events previously detected by spectroscopy experiments (which typically happens in femtoseconds). The charged defects are also responsible for the abnormally softened phonons at X point (Fig. 3a). As shown in Extended Data Fig. 9, the mode at X point is a collective vibration of two separate and interpenetrated hydrogen bond networks. At high densities, it is the only mode that is completely dominated by the nonbonding repulsion between O atoms. Since the diffusion of the charged defects is at a similar time scale to the O lattice vibration, their charge distribution can respond to the phonon vibrations concurrently. As a result, the electrostatic attraction between charged defects in the two networks softens the O-O repulsion, leading to the anomaly around the X point. Here thermal transport in the O sublattice serves as a powerful tool to probe and reveal the global motions of the H sublattice over the corresponding time and spatial scales.

Our study is the first to demonstrate the collective quantum motion of protons occurring in a high-pressure, high-density phase of ice. The collective tunneling under high pressure yields a new perspective to explore the ice VII-X phase transition. Moreover, the analysis of coupling between the O phonons and proton tunneling deepens our understanding of the scattering mechanism between acoustic phonons and atomic tunneling that is only scarcely demonstrated previously. The thermal conductivity of ice we measured over a wider range of pressures also provides an important benchmark in the study of the evolution and internal dynamics of icy planets.

## Main References

1. Nellis, W. J. *et al.* The Nature of the Interior of Uranus Based on Studies of Planetary Ices at High Dynamic Pressure. *Science* **240**, 779–781 (1988).

2. Cavazzoni, C. *et al.* Superionic and Metallic States of Water and Ammonia at Giant Planet Conditions. *Science* **283**, 44–46 (1999).

3. Bina, C. R. & Navrotsky, A. Possible presence of high-pressure ice in cold subducting slabs. *Nature* **408**, 844–847 (2000).

4. Hubbard, W. B. Neptune's Deep Chemistry. *Science* **275**, 1279–1280 (1997).

5. Sun, B. *et al.* High frequency atomic tunneling yields ultralow and glass-like thermal conductivity in chalcogenide single crystals. *Nat. Commun.* **11**, 6039 (2020).

6. Chen, B., Hsieh, W.-P., Cahill, D. G., Trinkle, D. R. & Li, J. Thermal conductivity of compressed $H_2O$ to 22 GPa: A test of the Leibfried-Schlömann equation. *Phys. Rev. B* **83**, 132301 (2011).

7. Lin, L., Morrone, J. A. & Car, R. Correlated Tunneling in Hydrogen Bonds. *J. Stat. Phys.* **145**, 365–384 (2011).

8.  Morrone, J. A., Lin, L. & Car, R. Tunneling and delocalization effects in hydrogen bonded systems: A study in position and momentum space. *J. Chem. Phys.* **130**, 204511 (2009).

9.  Zhang, L., Wang, H., Car, R. & E, W. Phase Diagram of a Deep Potential Water Model. *Phys. Rev. Lett.* **126**, 236001 (2021).

10. Drechsel-Grau, C. & Marx, D. Quantum Simulation of Collective Proton Tunneling in Hexagonal Ice Crystals. *Phys. Rev. Lett.* **112**, 148302 (2014).

11. Merolle, M., Garrahan, J. P. & Chandler, D. Space–time thermodynamics of the glass transition. *Proc. Natl. Acad. Sci. U. S. A.* **102**, 10837–10840 (2005).

12. Ye, Q.-J., Zhuang, L. & Li, X.-Z. Dynamic Nature of High-Pressure Ice VII. *Phys. Rev. Lett.* **126**, 185501 (2021).

13. Umemoto, K. *et al.* Nature of the Volume Isotope Effect in Ice. *Phys. Rev. Lett.* **115**, 173005 (2015).

14. Zha, C.-S., Tse, J. S. & Bassett, W. A. New Raman measurements for $H_2O$ ice VII in the range of 300 $cm^{-1}$ to 4000 $cm^{-1}$ at pressures up to 120 GPa. *J. Chem. Phys.* **145**, 124315 (2016).

15. Noguchi, N. & Okuchi, T. Self-diffusion of protons in $H_2O$ ice VII at high pressures: Anomaly around 10 GPa. *J. Chem. Phys.* **144**, 234503 (2016).

16. Komatsu, K. *et al.* Anomalous hydrogen dynamics of the ice VII–VIII transition revealed by high-pressure neutron diffraction. *Proc. Natl. Acad. Sci. U. S. A.* **117**, 6356–6361 (2020).

17. Goncharov, A. F., Struzhkin, V. V., Somayazulu, M. S., Hemley, R. J. & Mao, H. K.

Compression of Ice to 210 Gigapascals: Infrared Evidence for a Symmetric Hydrogen-Bonded Phase. *Science* **273**, 218–220 (1996).

18. Meier, T., Petitgirard, S., Khandarkhaeva, S. & Dubrovinsky, L. Observation of nuclear quantum effects and hydrogen bond symmetrisation in high pressure ice. *Nat. Commun.* **9**, 2766 (2018).

19. Cahill, D. G. Analysis of heat flow in layered structures for time-domain thermoreflectance. *Rev. Sci. Instrum.* **75**, 5119–5122 (2004).

20. Okada, T., Iitaka, T., Yagi, T. & Aoki, K. Electrical conductivity of ice VII. *Sci. Rep.* **4**, 5778 (2014).

21. Liu, B., Gao, Y., Han, Y., Ma, Y. & Gao, C. In situ electrical conductivity measurements of $H_2O$ under static pressure up to 28 GPa. *Phys. Lett. A* **380**, 2979–2983 (2016).

22. Wang, H., Zhang, L., Han, J. & E, W. DeePMD-kit: A deep learning package for many-body potential energy representation and molecular dynamics. *Comput. Phys. Commun* **228**, 178–184 (2018).

23. Habershon, S., Manolopoulos, D. E., Markland, T. E. & Miller, T. F. Ring-Polymer Molecular Dynamics: Quantum Effects in Chemical Dynamics from Classical Trajectories in an Extended Phase Space. *Annu. Rev. Phys. Chem.* **64**, 387–413 (2013).

24. Luo, R. & Yu, K. Capturing the nuclear quantum effects in molecular dynamics for lattice thermal conductivity calculations: Using ice as example. *J. Chem. Phys.* **153**, 194105 (2020).

25. Kong, L. T., Bartels, G., Campañá, C., Denniston, C. & Müser, M. H. Implementation of Green's function molecular dynamics: An extension to LAMMPS. *Comput. Phys. Commun.* **180**, 1004–1010 (2009).

26. Kong, L. T. Phonon dispersion measured directly from molecular dynamics simulations. *Comput. Phys. Commun.* **182**, 2201–2207 (2011).

27. Trybel, F., Cosacchi, M., Meier, T., Axt, V. M. & Steinle-Neumann, G. Proton dynamics in high-pressure ice-VII from density functional theory. *Phys. Rev. B* **102**, 184310 (2020).

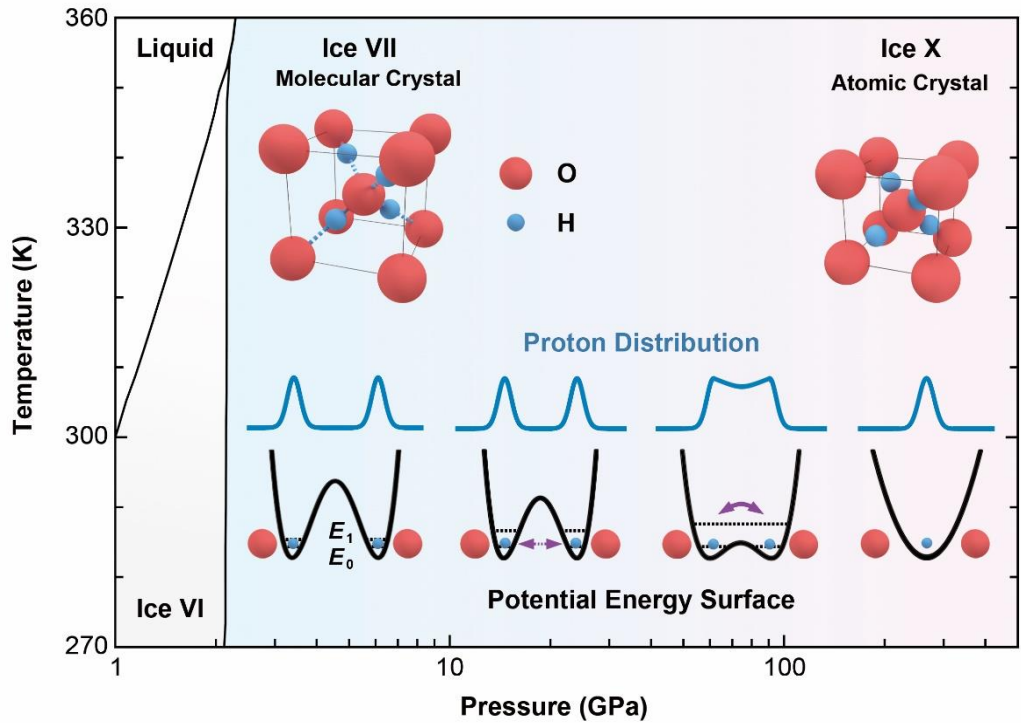

**Fig. 1 | Phase diagram of $H_2O$ and schematic of VII-X phase transition with proton symmetrization.** The phase diagram is from a past report[9] and the boundary between ice VII and X remains controversial. Inset shows the structure of ice VII and ice X, as well as the potential energy of a proton situated between two oxygen atoms and proton distribution probability. $E_0$ and $E_1$ (black dashed lines) are the zero-point energy and first excited state of proton, and their difference represents the tunnel splitting.

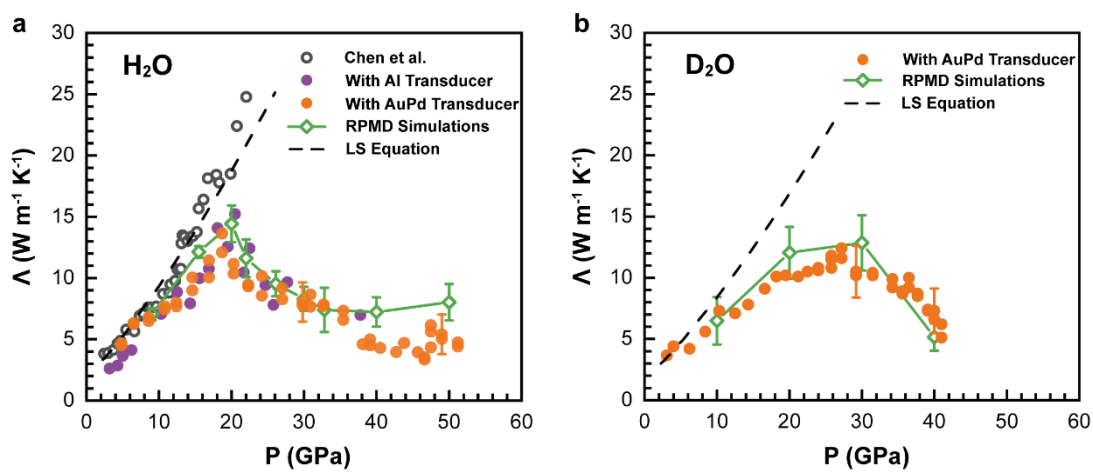

**Fig. 2 | Thermal conductivity of ice under hydrostatic pressure. a, b,** Thermal conductivity of $H_2O$ and $D_2O$. The purple and orange circles are TDTR results in this study with Al and AuPd transducer, respectively. The green line is results of RPMD simulations. The black dash line is calculated with the classical LS equation[6]. The details of calculating measurement uncertainties are in the Methods.

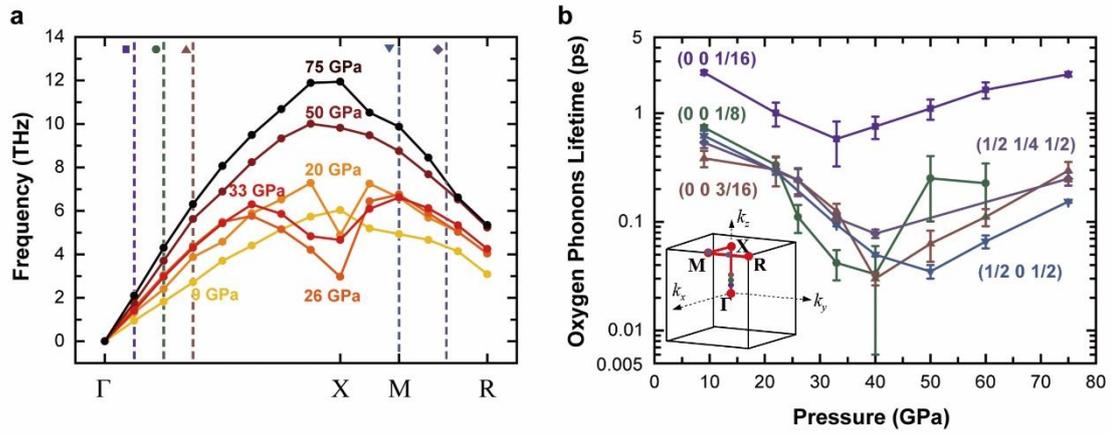

**Fig. 3 | Phonon properties of oxygen sublattice in $H_2O$ under pressure. a**, Pressure dependence of phonon dispersion. We only show a branch of acoustic mode for clarity. **b**, Pressure dependence of typical phonons lifetime. The positions of the phonons in the reciprocal space are also labeled in (**a**) with dashed lines. The inset is the first Brillouin zone of the oxygen sublattice.

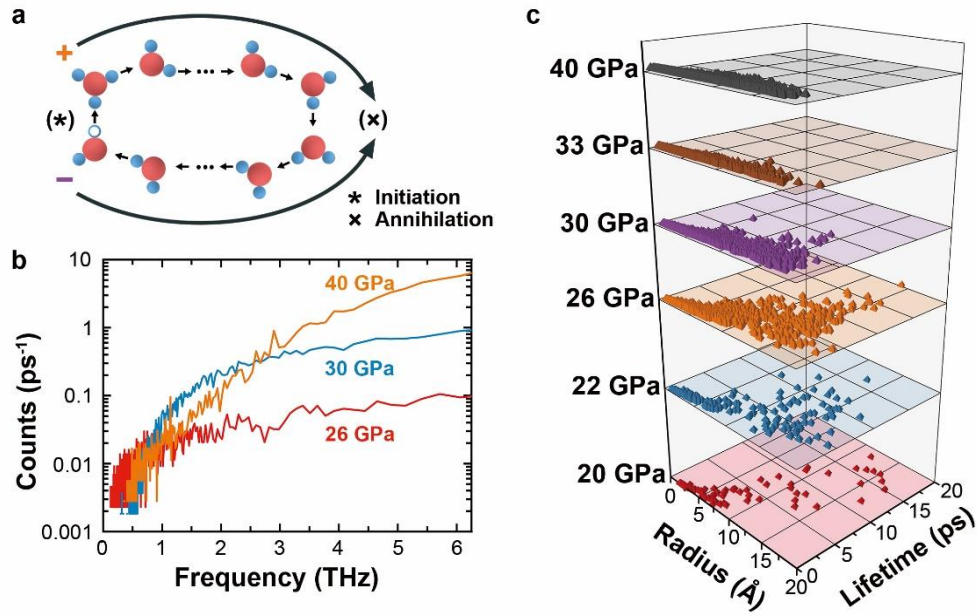

**Fig. 4 | Statistics of proton tunneling events at high pressures. a**, Schematic illustration of a collective tunneling loop. The tunneling of a proton occurs by excitation of charge defects (a pair of $H_3O^+$ and $OH^-$). The charge defect pair exhibits domino-like migrations along the look with more proton tunneling events until being annihilated at the other side of the loop. We define radius of the collective tunneling loop to be the geometrically effective radius of the tunneling path and lifetime of the collective tunneling loop to be the time between the excitation and the annihilation of the charge defects. **b**, Statistics of tunneling frequency at different pressures. The tunneling frequency is the inversion of the lifetime. We only show tunneling frequency less than $k_B T$ with $T$ of 300 K, where most heat-carrying phonons lie. **c,** Pressure dependence of the collective tunneling loops statistics.

## Methods

**TDTR measurement** Liquid $H_2O$ (HPLC grade; Macklin) or $D_2O$ (99.96%; Aladdin) was loaded in a symmetric DAC with 300-µm-culet diamonds and a steel gasket pre-indented to 40 µm with a 100 µm aperture. An ~15 µm thick sheet of muscovite mica (grade V-1; SPI Supplies) coated with 90-nm-thick Al or $Au_{0.95}Pd_{0.05}$ transducer was put on the culet of diamond to prevent heat from flowing into the diamond. The metal films create a temperature rise after a train of pump laser pulses with 6 µm or 12 µm $1/e^2$ diameter of Gaussian beam and ~785nm wavelength travel through the diamond and Ice VII and finally are absorbed by metal films. A probe pulse delayed in time with respect to the pump carries the signal of transient temperature decay. With the pressure increases, the phase transition to ice VII was confirmed by Raman spectroscopy (Extended Data Fig. 10). With a 10.1 MHz modulation frequency, thermal effusivity $\sqrt{\Lambda C_V}$ of Ice VII was directly obtained (Extended Data Fig. 3) by comparing the bidirectional thermal diffusion model (Details are shown in Supplementary Information) and the ratio of the in-phase ($V_{in}$) and out-of-phase ($V_{out}$) thermoreflectance signals with different delayed time measured by lock-in amplifier[19].

**Pressure calibration** Ruby spheres were loaded in the DAC sample chamber for pressure calibration. A 532 nm continuous wave laser with 0.5 mW power was used to excite the ruby fluorescence (Extended Data Fig. 1a). The wavelength of ruby R1 line shift with pressure was collected by a spectrometer (Ocean Optics HR4000) with 0.06 nm resolution (Extended Data Fig. 11). The pressure formula is

$$P = \frac{A}{B}\left[\left(\frac{\lambda}{\lambda_0}\right)^B - 1\right]$$

where $\lambda$ is the ruby R1 line at high pressure, $\lambda_0$ is the ruby R1 line (694.24 nm) at ambient pressure, $A = 1920$ GPa, and $B = 9.61$[28]. The $\lambda$ was obtained by fitting the fluorescence data with the sum of two Voigt functions[29]. The pressure calibration also can be verified by the Brillouin scattering data of Ice VII (Extended Data Fig. 4). We estimate ± 0.5 GPa uncertainty considering the spectrometer resolution.

**Pressure-dependent TDTR thermal parameters** The pressure-dependent parameters include $\Lambda$ of each layer, $C_V$ of each layer, and the thickness of the metal transducers.

The cross-plane $\Lambda$ of muscovite mica at high pressure has been measured (Extended Data Fig. 12), which is consistent with previous study[30,31]. The TDTR measurement is not sensitive to the in-plane $\Lambda$ of mica, which was calculated by Leibfried-Schlömann Equation based on the in-plane elastic constant $C_{11}$[30]. The thermal conductivities of metal transducers were fixed at values (93 W m$^{-1}$ K$^{-1}$ for AuPd and 187 W m$^{-1}$ K$^{-1}$ for Al) derived from the electrical conductivity at ambient pressure based on the Wiedemann–Franz Law because the thermal conductivity of metal is not important after 300 ps when the heat has already flowed through the metal layer in TDTR. The thermal conductivity of ice was obtained by fitting the TDTR data after 300 ps with the bidirectional thermal diffusion model.

The volumetric heat capacity of mica under pressure was from previous reports[30,31], which used the temperature dependence of the heat capacity combined with pressure dependence of the atomic density and elastic constants. The volumetric heat capacity of Al and AuPd was from a past report[32] where the heat capacity is the function of Debye temperature and atomic density based on the Debye model. The Debye temperature is assumed $T_{\text{Debye}} \propto K_T^{1/2} N^{-1/6}$, where $K_T$ is the isothermal bulk modulus and $N$ is the atomic density. Both $K_T$ and $N$ can be calculated from the pressure-volume equations of state (EOS) for Al[33] and Au[34]. The volumetric heat capacity of ice is calculated by quantum MD (Extended Data Fig. 5).

The thickness of metal transducers at ambient pressure was measured by picosecond acoustics:

$$h_{metal} = \frac{1}{2} v \Delta t_{echo}$$

Where $v$ is the longitudinal speed of sound (6.42 nm/ps for Al and 3.3 nm/ps for AuPd[32]) and $\Delta t_{echo}$ is the interval time for an acoustic signal that usually exists before 100 ps in TDTR. The thickness of metal transducers at high pressure was calculated by EOS of Al and Au combined with an assumption that the metal layer plastically deforms to accommodate the mismatch of the mica substrate, which is the same method as the previous literature[30]. Considering that there is no reported data about the EOS of AuPd, we assume the 5% Pd content has limited change for the EOS of Au. The two principal axes in the in-plane of mica decrease with pressure where the lattice constants of muscovite mica under pressure are taken from the past report[35]. The two principal axes of metals paralleled to those of mica decrease by the same percentage. The other principal axis of metals along the thickness direction can be calculated based on the EOS of metals.

The parameters used in the bidirectional thermal diffusion model are listed in Extended Data Table 1.

**Uncertainty analysis** To estimate the uncertainty of TDTR measurements, sensitivity:

$$S_\alpha = \frac{\partial \ln\left(-\frac{V_{in}}{V_{out}}\right)}{\partial \ln \alpha}$$

for the parameters $\alpha$ is calculated based on the thermal diffusion model. The parameters include $\Lambda$ and $C_V$ of each layer, the thickness of metal transducer, the spot size of the laser beam, and phase signal of the lock-in amplifier. The uncertainty of thermal conductivity of ice VII is:

$$\left(\frac{\Delta \Lambda}{\Lambda}\right)^2 = \sum_\alpha \left(\frac{S_\alpha}{S_\Lambda} \frac{\Delta \alpha}{\alpha}\right)^2$$

We found that uncertainties in all the parameters propagate to ~20% measurement uncertainty on thermal conductivity below 30 GPa, and ~30% measurement uncertainty at 50 GPa.

# Molecular dynamics (MD) simulations

## A. Construction of Potential Energy Surface

The accuracy of the potential energy surface (PES) is very important to ensure the fidelity of our simulation. The conventional nonreactive water force fields are not capable to describe the tunneling of protons, while ab initio molecular dynamics (AIMD) is too expensive for large-scale simulations that are required in thermal transport studies. Therefore, a reactive machine learning (ML) PES was generated using DeepMD-kit[9,36] with reference energies and forces computed using Perdew–Burke-Ernzerhof (PBE) density functional. To ensure the quality of the PES in quantum MD simulations, we train the ML potential following this procedure:

1. Classical AIMD simulations were conducted within a wide range of temperatures (from 78 K to 500 K) and pressures (from 5 GPa to 100 GPa) using a 64-molecule ice box. In total, 32 trajectories at given pressure were generated with NVT runs, with each trajectory run for 20 ps. Then 500 configurations were drawn uniformly from all trajectories and their energies and forces were computed, giving the initial training dataset.

2. It is noted that due to the high energy barrier of proton transfer, the transition region is severely under-sampled at low pressures. To solve this problem, we first drew 500 configurations from high pressure (>75 GPa) simulations, in which protons were fluctuating around symmetrized positions. Then these high-pressure configurations were simply rescaled to the simulated densities at lower pressures and were added to the training set.

3. A ML model ($M_0$) was trained using the available PBE energy and force data.

4. 20 ps path-integral MD (PIMD) simulations with 32 beads were performed at 500 K and from 9-100 GPa, using $M_0$ and the 64-molecule ice box. Then 1280 bead configurations (40 frames) were drawn from each PIMD trajectory. The PBE energies and forces of the PIMD samples were then computed and compared with the values predicted by $M_0$. Since $M_0$ were trained using classical MD data, it could not describe some of the PIMD geometries correctly. All configurations with a force error larger than 1.5 meV/A were selected and added to the training set.

5. The ML model was retrained with the newly added PIMD data, and steps 4 and 5 were repeated until no more new data points can be generated.

With this procedure, we made sure that all the geometries relevant in the production PIMD runs can be correctly described by the resulting ML model.

### B. Details of Thermal Transport Simulations

Nuclear quantum effects (NQE) have a significant impact on both the thermodynamic and dynamic behaviors in ice VII. In the production run, ring-polymer molecular dynamics (RPMD) was used to propagate dynamics, and the centroid trajectories were utilized to compute the heat flux. Such a method was benchmarked against quantum Boltzmann transport equation (BTE) results in our previous studies, and was proved to be reliable in ice system[24]. The procedure is described as following:

All production runs were performed using i-PI[37], interfaced with the DeepPotential energy and force calculator. For all systems (including both ice VII and X), we used the 4×4×16 supercell as our simulation box, containing 3072 atoms. To obtain the thermal conductivity, a long NVT PIMD simulation (more than 500 ps) was run first, using the path-integral Langevin equation thermostat (PILE)[38] and a timestep of 0.5 fs. Then, no less than 64 structures were randomly selected from the NVT trajectory and used as the initial configurations for the following simulations. Then RPMD was run in the NVE ensemble to generate real-time dynamics. Each NVE trajectory was propagated for 20-40 ps with a timestep of 0.2 fs, depending on the convergence speed of the heat flux time correlation functions (TCFs) at different conditions. The heat flux TCFs $c_{JJ}(t)$ were then computed using the centroid positions and velocities:

$$\vec{J}(\{\vec{r}_a\},\{\vec{v}_a\}) = \sum_a e_a \vec{v}_a + \sum_a \mathbf{w}_a \cdot \vec{v}_a.$$

In here, $e_a$ is the total atomic energy:

$$e_a = \frac{1}{2} m_a v_a^2 + U_a.$$

The atomic potential energy $U_a$ is naturally given by the DeepPotential model as it assumes: $U = \sum_a U_a$. Correspondingly, the atomic virial tensor is computed as:

$$\mathbf{w}_a = \vec{r}_{ab} \otimes \frac{\partial U_b}{\partial \vec{r}_a}.$$

Green-Kubo formula was then utilized to calculate the thermal conductivity:

$$\begin{aligned}\Lambda_{\alpha\beta} &= \frac{V}{k_B T^2} \int_0^\infty \langle J_\alpha(0) J_\beta(t) \rangle_{NVT} dt \\ &= \frac{V}{k_B T^2} \int_0^\infty c_{JJ}^{\alpha\beta}(t) dt,\end{aligned}$$

where $V$, $k_B$, and $T$ represent the volume, Boltzmann constant, and temperature, respectively, and $\alpha$, $\beta$ are the dimension indices. To understand the result quality, we also compare our result via direct Green-Kubo integration with cepstral analysis methods[39]. The deviation of our results is consistent with that of cepstral analysis and therefore the convergence of our results is guaranteed.

The vibrational eigenvectors and the dispersion relationship of the oxygen sublattice phonons were extracted using the Green Function method[26,40]. Then, following the recipe given by McGaughey et al.[41], the oxygen displacement ($\vec{u}$) and velocity ($\dot{\vec{u}}$) trajectories were projected onto the phonon eigenvectors ($\vec{e}$) to obtain the dynamics of the corresponding phonon modes:

$$\begin{cases} x(\lambda,\vec{k};t) = \sum_{b,l}\left(\frac{m_b}{N}\right)^{1/2}\exp\left(i\vec{k}\cdot\vec{r}_0^l\right)\left[\vec{e}_b^*(\lambda,\vec{k})\cdot\vec{u}(b,l;t)\right] \\ \dot{x}(\lambda,\vec{k};t) = \sum_{b,l}\left(\frac{m_b}{N}\right)^{1/2}\exp\left(i\vec{k}\cdot\vec{r}_0^l\right)\left[\vec{e}_b^*(\lambda,\vec{k})\cdot\dot{\vec{u}}(b,l;t)\right]. \end{cases}$$

Here, $b$ is the oxygen index in the unit-cell and $l$ is the unit-cell index inside the MD supercell. $\vec{r}_0^l$ denotes the real space shift of the $l$th unit-cell and $\lambda$ denotes different phonon bands. The phonon potential, kinetic, and total energies were computed as:

$$U(\lambda,\vec{k};t) = \frac{1}{2}\omega^2(\lambda,\vec{k})|x|^2$$
$$T(\lambda,\vec{k};t) = \frac{1}{2}|\dot{x}|^2$$
$$E = T + U.$$

The integration of the phonon energy TCF thus gives the phonon lifetime,

$$\tau_{\lambda,k} = \int_0^\infty \frac{<E_{\lambda,k}(t)E_{\lambda,k}(0)>}{<E_{\lambda,k}(0)E_{\lambda,k}(0)>}dt.$$

Constant volume heat capacity ($C_V$) was calculated following the protocol given in reference[42]. The heat capacity is equivalent to the energy fluctuation in thermodynamics,

$$C_v(\beta) = k_B\beta^2\left\{\frac{1}{Z(\beta)}\frac{\partial^2 Z(\beta)}{\partial \beta^2} - [\frac{1}{Z(\beta)}\frac{\partial Z(\beta)}{\partial \beta}]^2\right\}.$$

In practice, we evaluated it by the double virial estimators implemented in i-PI, which is defined as,

$$C_v(\beta) = k_B\beta^2\{<\varepsilon_v^2> - <\varepsilon_v>^2 - <\varepsilon_v'>\},$$

where $\beta = \frac{1}{k_B T}$. $\varepsilon_v$ is internal energy based on coordinate scaling and $\varepsilon_v^{'}$ is its derivative with respect to $\beta$.

**Methods references**

28. Dewaele, A., Torrent, M., Loubeyre, P. & Mezouar, M. Compression curves of transition metals in the Mbar range: Experiments and projector augmented-wave calculations. *Phys. Rev. B* **78**, 104102 (2008).

29. Raghavan, S., Imbrie, P. K. & Crossley, W. A. Spectral Analysis of R-lines and Vibronic Sidebands in the Emission Spectrum of Ruby Using Genetic Algorithms. *Appl. Spectrosc.* **62**, 759–765 (2008).

30. Hsieh, W.-P., Chen, B., Li, J., Keblinski, P. & Cahill, D. G. Pressure tuning of the thermal conductivity of the layered muscovite crystal. *Phys. Rev. B* **80**, 180302 (2009).

31. Shieh, S. R., Hsieh, W.-P., Tsao, Y.-C., Crisostomo, C. & Hsu, H. Low Thermal Conductivity of Carbon Dioxide at High Pressure: Implications for Icy Planetary Interiors. *J. Geophys. Res. Planets* **127**, (2022).

32. Hohensee, G. T., Wilson, R. B. & Cahill, D. G. Thermal conductance of metal-diamond interfaces at high pressure. *Nat. Commun.* **6**, 6578 (2015).

33. Greene, R. G., Luo, H. & Ruoff, A. L. Al as a Simple Solid: High Pressure Study to 220 GPa (2.2 Mbar). *Phys. Rev. Lett.* **73**, 2075–2078 (1994).

34. Takemura, K. & Dewaele, A. Isothermal equation of state for gold with a He-pressure medium. *Phys. Rev. B* **78**, 104119 (2008).

35. Curetti, N., Levy, D., Pavese, A. & Ivaldi, G. Elastic properties and stability of coexisting 3T and 2M$_1$ phengite polytypes. *Phys. Chem. Miner.* **32**, 670–678 (2006).

36. Zhang, L. *et al.* End-to-end Symmetry Preserving Inter-atomic Potential Energy Model for Finite and Extended Systems. *Adv. Neural Inf. Process. Syst.* **31**, (2018).

37. Kapil, V. *et al.* i-PI 2.0: A universal force engine for advanced molecular simulations. *Comput. Phys. Commun.* **236**, 214–223 (2019).

38. Ceriotti, M., Parrinello, M., Markland, T. E. & Manolopoulos, D. E. Efficient stochastic thermostatting of path integral molecular dynamics. *J. Chem. Phys.* **133**, 124104 (2010).

39. Ercole, L., Marcolongo, A. & Baroni, S. Accurate thermal conductivities from optimally short molecular dynamics simulations. *Sci. Rep.* **7**, 15835 (2017).

40. Campañá, C. & Müser, M. H. Practical Green's function approach to the simulation of elastic semi-infinite solids. *Phys. Rev. B* **74**, 075420 (2006).

41. McGaughey, A. J. H. & Larkin, J. M. Predicting phonon properties from equilibrium molecular dynamics simulations. *Annu. Rev. Heat Transf.* **17**, 49–87 (2014).

42. Yamamoto, T. M. Path-integral virial estimator based on the scaling of fluctuation coordinates: Application to quantum clusters with fourth-order propagators. *J. Chem. Phys.* **123**, 104101 (2005).

43. Polian, A. & Grimsditch, M. New High-Pressure Phase of H$_2$O: Ice X. *Phys. Rev. Lett.* **52**, 1312–1314 (1984).

44. Ahart, M. *et al.* Brillouin scattering of H$_2$O ice to megabar pressures. *J. Chem. Phys.*

**134**, 124517 (2011).

45. Grande, Z. M. *et al.* Pressure-driven symmetry transitions in dense $H_2O$ ice. *Phys. Rev. B* **105**, 104109 (2022).

**Acknowledgements** The authors would like to thank Dr. Fang Liu and Prof. Xinqiang Wang from Peking University for depositing AuPd films. B. S. acknowledges support from National Science Foundation of China under Grant No. 12004211, NSFC-ISF Joint Scientific Research Program under Grant No. 52161145502, Shenzhen Science and Technology Program grants RCYX20200714114643187 and WDZC20200821100123001, Tsinghua Shenzhen International Graduate School grants QD2021008N and JC2021008. K. Y. acknowledges support from National Natural Science Foundation of China under Grant No. 22103048 and Tsinghua Shenzhen International Graduate School under Grant No. HW2020009. J. W. acknowledges support from The U.S. Department of Energy, Office of Science, Office of Basic Energy Sciences, Materials Sciences and Engineering Division under Contract No. DE-AC02-05-CH11231 (EMAT program KC1201).

**Author contributions** B.S. and K.Y. conceived of the concept. Y.W. planned and completed the experiments. R.L. performed the simulations. J.C., X.Z., and S.W. provided the support on high-pressure technique. Y.W., R.L., J.W., F.K., K.Y., and B.S. wrote the manuscript. B.S. and K.Y. supervised the project.

**Competing interest declaration** Authors declare that they have no competing interests.

**Additional information**

Supplementary Information is available for this paper.

Correspondence and requests for materials should be addressed to Bo Sun and Kuang Yu.

**Extended Data Table 1. The parameters used in the bidirectional thermal diffusion model.**

| Pressure<br>Parameters | 10.8 GPa | 20.3 GPa | 29.8 GPa | 40.5 GPa | 49.9 GPa |
|---|---|---|---|---|---|
| $C_V$ of mica<br>(J cm$^{-3}$ K$^{-1}$) | 2.50 | 2.65 | 2.79 | 2.94 | 3.05 |
| $\Lambda_{cross-plane}$ of mica<br>(W m$^{-1}$ K$^{-1}$) | 1.94 | 3.73 | 5.79 | 8.13 | 10.03 |
| $\Lambda_{in-plane}$ of mica<br>(W m$^{-1}$ K$^{-1}$) | 5.31 | 6.48 | 7.85 | 9.72 | 13.15 |
| $C_V$ of Al<br>(J cm$^{-3}$ K$^{-1}$) | 2.61 | 2.72 | 2.79 | 2.84 | 2.88 |
| $\dfrac{h_{high\ pressure}}{h_{ambient\ pressure}}$ of Al | 0.94 | 0.91 | 0.89 | 0.87 | 0.86 |
| $C_V$ of AuPd<br>(J cm$^{-3}$ K$^{-1}$) | 2.59 | 2.67 | 2.75 | 2.83 | 2.89 |
| $\dfrac{h_{high\ pressure}}{h_{ambient\ pressure}}$ of AuPd | 0.997 | 0.992 | 0.988 | 0.983 | 0.979 |
| $C_V$ of H$_2$O<br>(J cm$^{-3}$ K$^{-1}$) | 3.94 | 4.01 | 5.18 | 3.96 | 3.97 |
| $C_V$ of D$_2$O<br>(J cm$^{-3}$ K$^{-1}$) | 4.30 | 4.28 | 4.07 | 4.18 | - |

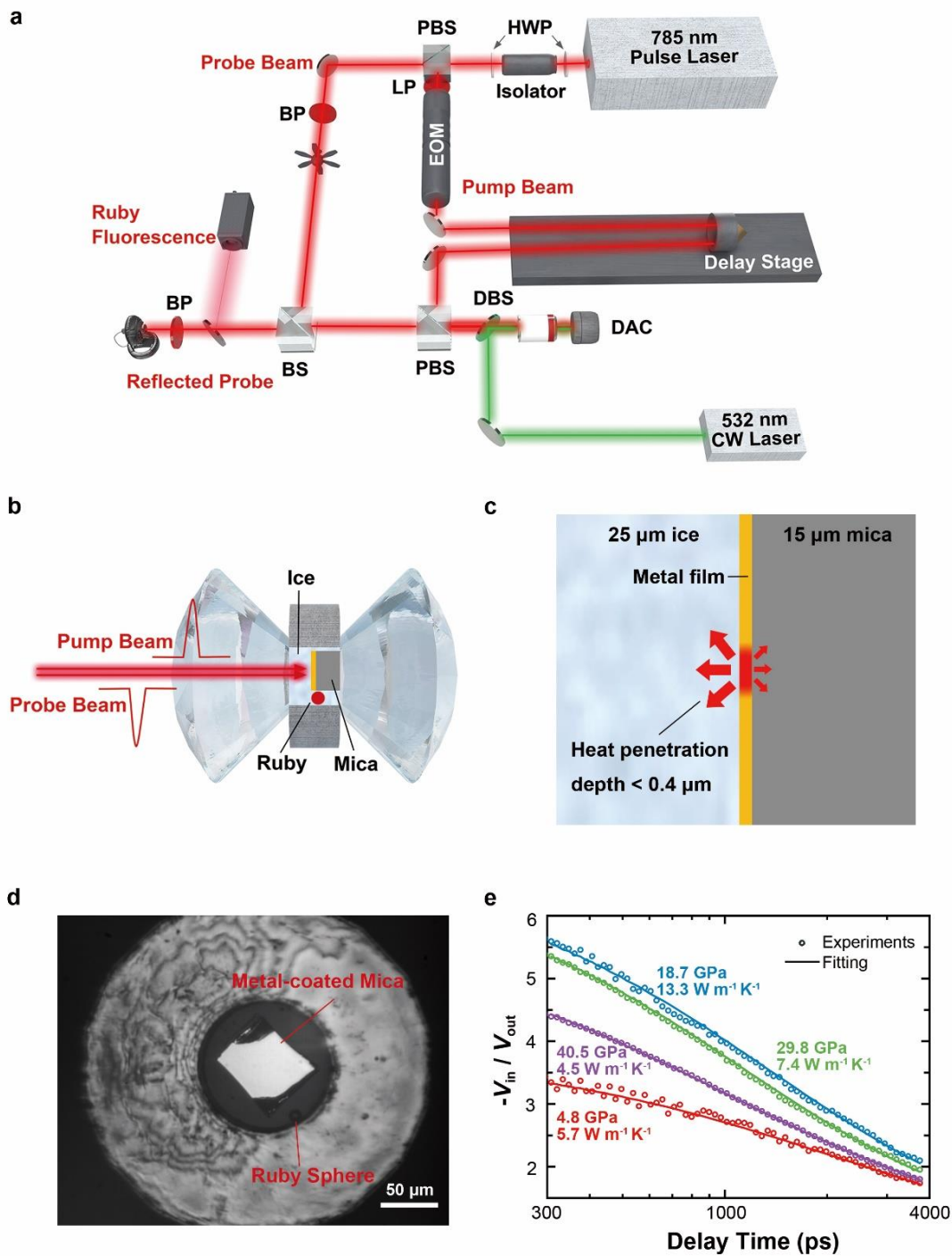

**Extended Data Fig. 1 | Schematic of TDTR setup and samples in a DAC. a**, Schematic of the TDTR setup. The element labeled HWP is a half-wave plate, PBS is a polarizing beam splitter, LP is a long-pass filter, BP is a band-pass filter, EOM is an electro-optic modulator, BS is a non-polarizing beam-splitter, DBS is a dichroic beam-splitter. **b**, A sheet of muscovite mica with 90 nm Al or AuPd transducer

thermally insulates the ice from the diamond in a DAC. The ruby sphere is used for pressure calibration. **c**, Schematic of heat dissipation process in TDTR **d**, Optical micrograph of metal-coated mica and ruby sphere in the DAC sample chamber from the top view. **e**, TDTR data (hollow circles) and fitted curves (lines) for $H_2O$ VII at different pressures.

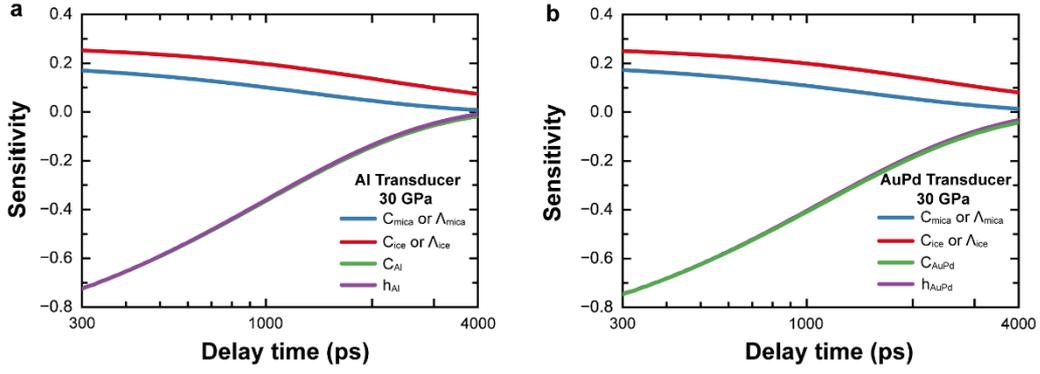

**Extended Data Fig. 2 | Sensitivity analysis of TDTR.** Sensitivity $S_\alpha$ to the parameters $\alpha$ used in the bidirectional thermal diffusion model of TDTR with **a**, Al and **b**, AuPd transducer at 30 GPa. $C_{mica}$, the volumetric heat capacity of muscovite mica; $\Lambda_{mica}$, the thermal conductivity of muscovite mica; $C_{ice}$, the volumetric heat capacity of ice VII; $\Lambda_{ice}$, the thermal conductivity of ice VII; $C_{Al/AuPd}$, the volumetric heat capacity of Al or AuPd; and $h_{Al/AuPd}$, the thickness of Al or AuPd. The $S_\alpha$ of other parameters are close to zero, thus not showed for clarity.

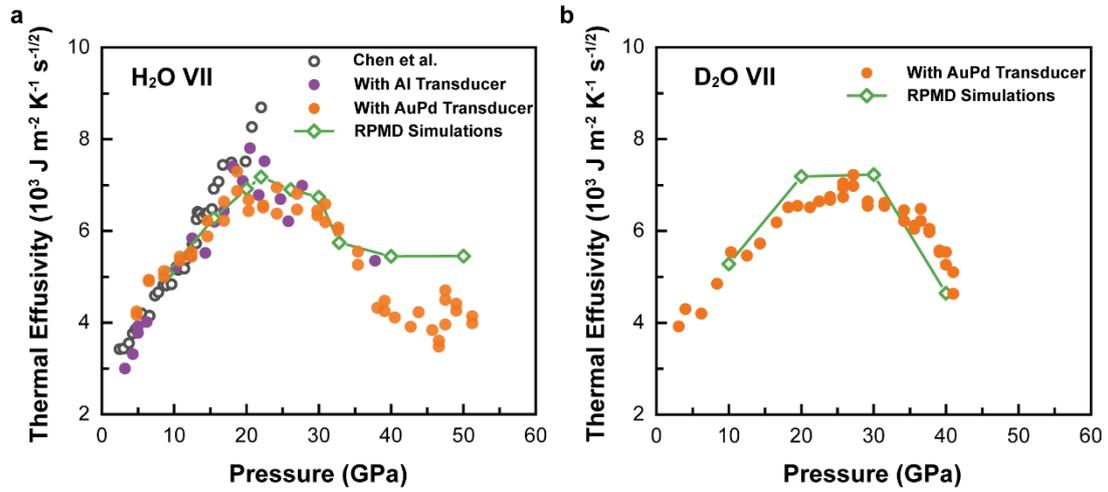

**Extended Data Fig. 3 | Thermal effusivity of ice VII at 300K. a, b,** Thermal effusivity of $H_2O$ VII and $D_2O$ VII. The purple and orange circles are TDTR results in this study with Al and AuPd transducer. The green line is the result of RPMD simulations.

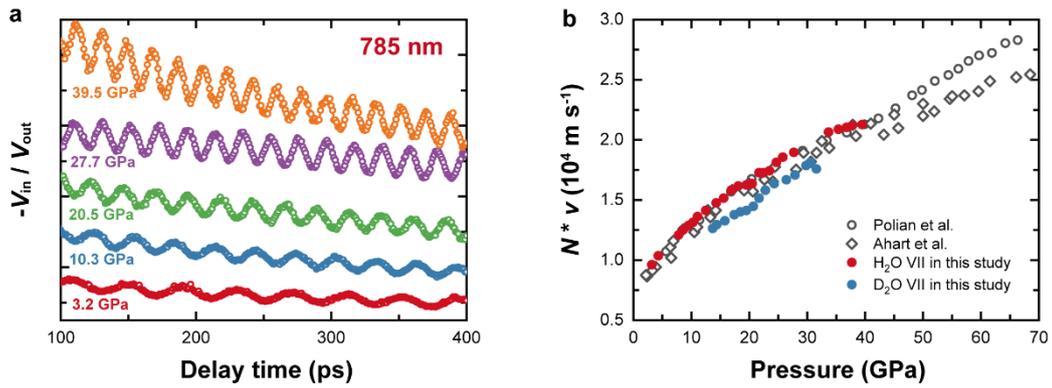

**Extended Data Fig. 4 | Time-domain stimulated Brillouin scattering of Ice VII. a**, Example data for the periodic oscillations in thermoreflectance signals that are used to in situ measure time-domain stimulated Brillouin scattering. **b**, Refraction index times speed of sound measured in ice VII (red circles for $H_2O$ and blue circles for $D_2O$). Past studies from Polian et al.[43] and Ahart et al.[44] are also plotted in hollow symbols. For Brillouin backscattering geometry, $f = 2Nv/\lambda$, where $f$ is the frequency shift of the scattered light, $N$ is the refraction index, $v$ is the longitudinal speed of sound, $\lambda$ is the laser wavelength. Considering this study used a different wavelength from previous studies, we used refraction index times speed of sound in (**b**) for comparison.

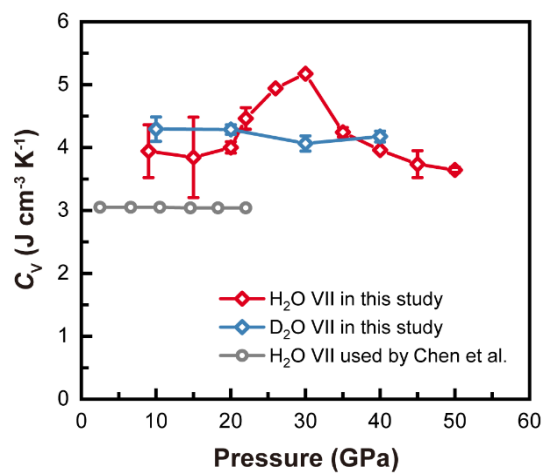

**Extended Data Fig. 5 | Volumetric heat capacity of $H_2O$ VII and $D_2O$ VII.** The colored lines are calculated by quantum MD in this study. Previous work by Chen et al.[6] approximated the $C_V$ of ice VII using the vibrational density of states of the fully ordered ice VIII, which are almost constant up to 20 GPa. The constant is reasonable because atomic density increases and thermally excited phonon modes decrease with pressure. In contrast, here we compute the $C_V$ of ice VII using quantum MD, explicitly accounting for both proton disorder and proton tunneling effects. It is noted that $C_V$ of ice VII anomalously increases above 20 GPa, which is another evidence that tunneling events markedly increase.

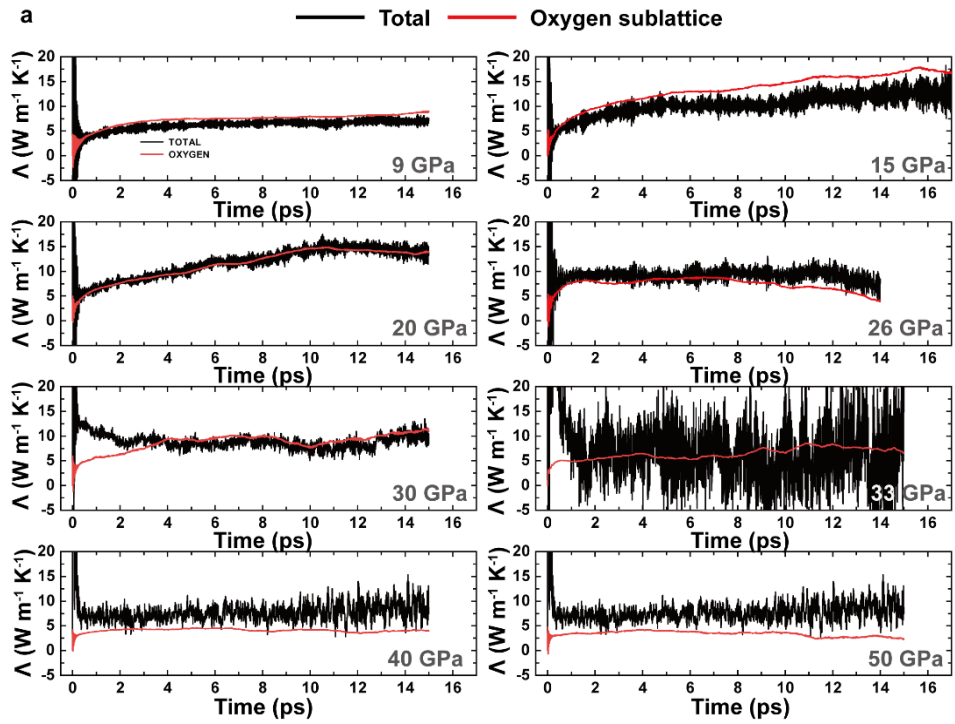

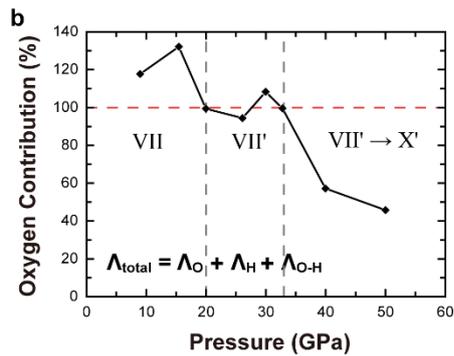

**Extended Data Fig. 6 | Thermal conductivity of H₂O VII and oxygen sublattice contribution calculated by RPMD. a**, Diagrams of time correlation functions integration at different pressures. The black and red lines are labeled as H$_2$O VII and oxygen sublattice. **b**, Contribution of oxygen sublattice to Λ extracted from (**a**).

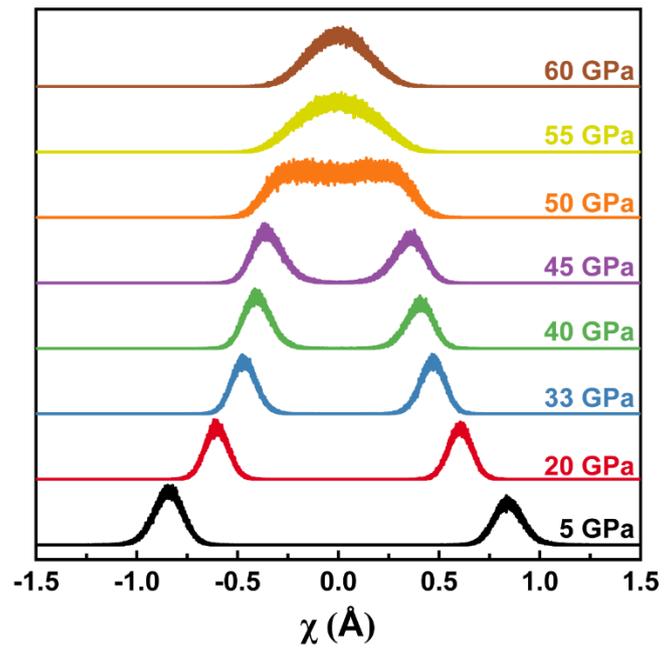

**Extended Data Fig. 7 | The proton distribution profile as a function of the proton position relative to the bond midpoint in ice VII.**

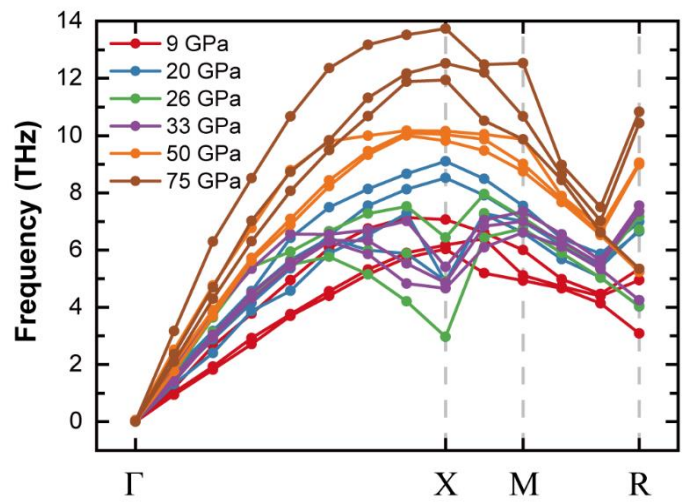

**Extended Data Fig. 8 | Pressure dependence of phonon dispersion of oxygen sublattice in H$_2$O.**

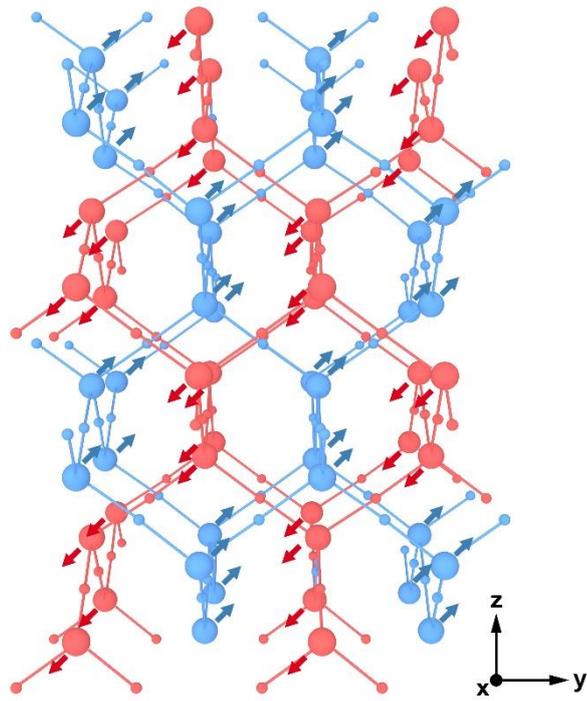

**Extended Data Fig. 9 | The schematic lattice vibration of oxygen sublattice phonon mode (0 0 1/2).**

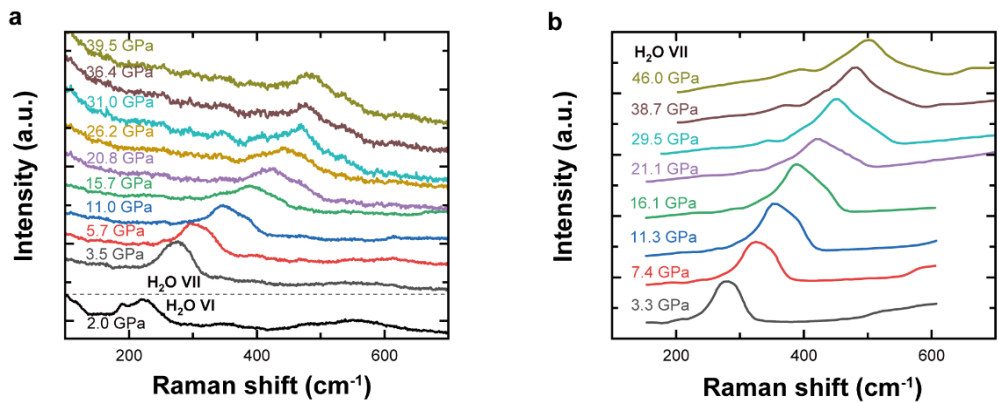

**Extended Data Fig. 10 | Raman spectroscopy of high-pressure $H_2O$. a**, Raman spectroscopy of $H_2O$ VI and VII in this study. The laser excitation of 638 nm with ~40 mW was used and the exposure time was 60 s repeated 10 times. **b,** Raman spectroscopy of $H_2O$ VII from the literature[45] for comparison.

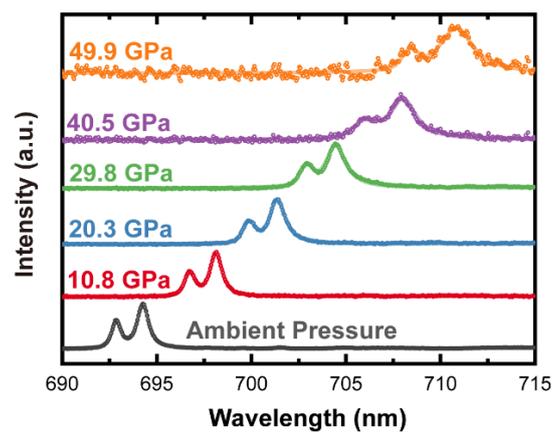

**Extended Data Fig. 11 | Ruby fluorescence for pressure calibration.**

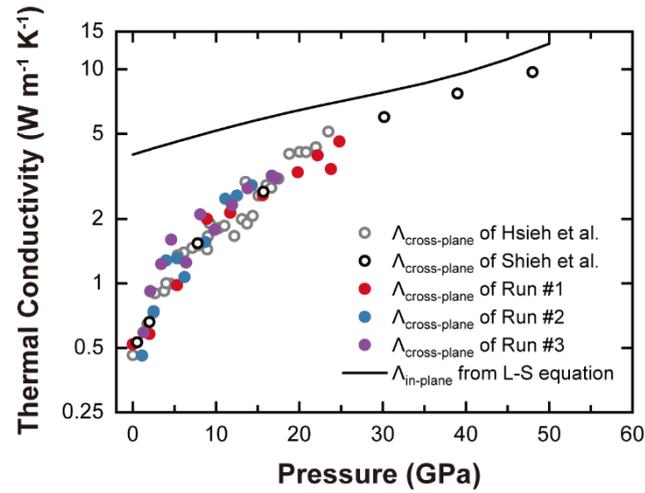

**Extended Data Fig. 12 | Thermal conductivity of muscovite mica.** The grey and black hollow circles are cross-plane Λ of muscovite mica reported by Hsieh et al.[30] and Shieh et al.[31]. The colored squares are measured cross-plane Λ of muscovite mica at high pressure in this study. The in-plane Λ of muscovite mica is calculated by Leibfried-Schlömann Equation based on the in-plane elastic constant $C_{11}$[30].

# Supplementary Information for

# Proton Collective Quantum Tunneling Induces Anomalous Thermal Conductivity of Ice under Pressure

Yufeng Wang[1,2]*, Ripeng Luo[2]*, Jian Chen[3], Xuefeng Zhou[3], Shanmin Wang[3], Junqiao Wu[4,5], Feiyu Kang[1,2], Kuang Yu[1,2]†, Bo Sun[1,2]†

* These authors contributed equally to this work

† e-mails: sun.bo@sz.tsinghua.edu.cn, yu.kuang@sz.tsinghua.edu.cn

**This file includes:**

      Bidirectional thermal diffusion model
      Video on phase transition from ice VI to ice VII

**Bidirectional thermal diffusion model**

In TDTR, the temperature distribution is calculated based on a layered structure. In our experiments, the in-plane thermal properties are isotropic. We can begin with the heat diffusion equation:

$$\Lambda_z \frac{\partial^2 T(r,z,t)}{\partial z^2} + \Lambda_r \left( \frac{\partial^2 T(r,z,t)}{\partial r^2} + \frac{1}{r}\frac{\partial T(r,z,t)}{\partial r} \right) = \rho C \frac{\partial T(r,z,t)}{\partial t}$$

where $\Lambda_z$ is cross-plane thermal conductivity, $\Lambda_r$ is in-plane thermal conductivity, $T$ is the temperature, $\rho$ is the density, $C$ is the heat capacity. Applying the Fourier transform and Hankel transform to the heat diffusion equation, we can obtain:

$$\frac{\partial^2 T(k,z,\omega)}{\partial z^2} - \left(4\pi^2 k^2 \frac{\Lambda_r}{\Lambda_z} + \frac{i\omega}{D_z}\right) T(k,z,\omega) = 0$$

Where $D_z = \frac{\rho C}{\Lambda_z}$. The solution of the equation can be expressed as a sum of thermal waves traveling forward and backward along the $z$ axis:

$$T_j(k,z,\omega) = T_j^+ \exp(u_j z) + T_j^- \exp(-u_j z)$$

$$u_j = \left( 4\pi^2 k^2 \frac{\Lambda_{rj}}{\Lambda_{zj}} + \frac{i\omega}{D_{zj}} \right)^{\frac{1}{2}}$$

where z is the coordinate in the layer $j = 1, 2, \cdots, n \cdots, N$ from the top layer to the bottom layer, $T_j^+$ and $T_j^-$ are the complex constants of the thermal waves traveling forward and backward for the layer $j$. We convert the temperature into a two-dimensional vector **T**:

$$\mathbf{T} = \begin{bmatrix} T_j^+(z) \\ T_j^-(z) \end{bmatrix}$$

Here, we consider the metal layer n to be heated at the surface by a periodic point source of unit power at angular frequency $\omega$. In order to obtain the temperature of the metal layer n, the total propagator $G_{\text{total}}(k,\omega)$ of the layer n is divided into upward heat flow and downward heat flow to calculate based on transfer matrix method[1–3]:

$$\begin{pmatrix} A^+ \\ A^- \end{pmatrix}_{j+1} = \frac{1}{2\gamma_{j+1}} \begin{pmatrix} e^{u_{j+1}h_{j+1}} & 0 \\ 0 & e^{-u_{j+1}h_{j+1}} \end{pmatrix} \begin{pmatrix} \gamma_{j+1}+\gamma_j & \gamma_{j+1}-\gamma_j \\ \gamma_{j+1}-\gamma_j & \gamma_{j+1}+\gamma_j \end{pmatrix} \begin{pmatrix} A^+ \\ A^- \end{pmatrix}_j$$

where $\gamma_j = \Lambda_j u_j$ and $h_i$ is the thickness of the $j$th layer. In practical applications of TDTR, heat cannot reach the other side of the top and bottom layer for $j=1$ or $N$. Therefore, $A_1^- = 0$ and $A_1^+ = 1$ for the top layer. The interface thermal conductance $G$ is treated as a 1 nm layer with almost zero heat capacity and choosing $\Lambda_z$ such that $G = \frac{\Lambda_z}{d}$. The propagator of upward direction is calculated by iterating from $j=1$ to n−2:

$$G_{\text{up}} = \frac{1}{\gamma_{n-1}} \frac{A_{n-1}^+ + A_{n-1}^-}{A_{n-1}^+ - A_{n-1}^-}$$

For the downward heat flow,

$$\begin{pmatrix} B^+ \\ B^- \end{pmatrix}_{j-1} = \frac{1}{2\gamma_{j-1}} \begin{pmatrix} -e^{u_{j-1}h_{j-1}} & 0 \\ 0 & e^{u_{j-1}h_{j-1}} \end{pmatrix} \begin{pmatrix} \gamma_{j-1}+\gamma_j & \gamma_{j-1}-\gamma_j \\ \gamma_{j-1}-\gamma_j & \gamma_{j-1}+\gamma_j \end{pmatrix} \begin{pmatrix} B^+ \\ B^- \end{pmatrix}_j$$

For the bottom layer, $B_N^+ = 0$ and $B_N^- = 1$, the propagator is:

$$G_{\text{down}} = \frac{1}{\gamma_n} \frac{B_n^+ + B_n^-}{B_n^+ - B_n^-}$$

The total propagator is:

$$G_{\text{total}} = \frac{1}{\frac{1}{G_{\text{down}}} - \frac{1}{G_{\text{up}}}}$$

In TDTR, the metal layer is heated by a pump laser beam with a Gaussian distribution. The intensity of $p(r)$ is:

$$p(r) = \frac{2A}{\pi w_0^2} \exp(-\frac{2r^2}{w_0^2})$$

where $w_0$ is the $1/e^2$ radius of the pump beam and $A$ is the amplitude of heat absorbed by the metal layer. The Hankel transform of $p(r)$ is:

$$P(k) = A\exp(-\frac{\pi^2 k^2 w_0^2}{2})$$

The distribution of temperature at the surface $\theta(r)$ is the inverse Hankel transform of the product of $G(k)$ and $P(k)$

$$\theta(r) = 2\pi \int_0^\infty P(k)G(k)J_0(2\pi kr)k \, dk$$

In TDTR, the surface temperature signal is detected by a probe laser beam with the same Gaussian distribution of intensity. The probe beam measures a weighted average of $\theta(r)$

$$\Delta T = \frac{4}{w_0^2} \int_0^\infty \theta(r)\exp(-\frac{2r^2}{w_0^2})r \, dr$$
$$= 2\pi A \int_0^\infty G(k)\exp(-\pi^2 k^2 w_0^2) \, dk$$

## References

1. Feldman, A. Algorithm for solutions of the thermal diffusion equation in a stratified medium with a modulated heating source. *High Temp. - High Press.* **31**, 293–298 (1999).

2. Xie, X., Dennison, J. M., Shin, J., Diao, Z. & Cahill, D. G. Measurement of water vapor diffusion in nanoscale polymer films by frequency-domain probe beam deflection. *Rev. Sci. Instrum.* **89**, 104904 (2018).

3. Cahill, D. G. Analysis of heat flow in layered structures for time-domain thermoreflectance. *Rev. Sci. Instrum.* **75**, 5119–5122 (2004).

s